\documentclass[reprint,amsmath,amssymb,aps]{revtex4-1}
\usepackage{graphicx}% Include figure files
\usepackage{dcolumn}% Align table columns on decimal point
\usepackage{bm}% bold math
\usepackage{color}
\usepackage{epstopdf}
\begin{document}

\preprint{APS/123-QED}

\title{Exploring interacting bulk viscous model with decaying vacuum density}

\author{Vinita Khatri}
\email[]{vinitakhatri\_2k20phdam501@dtu.ac.in}
\author{C. P. Singh}

\email[]{cpsingh@dce.ac.in (Corresponding author)}
\affiliation{Department of Applied Mathematics, \\Delhi Technological University, Delhi-110042, India}
\author{Milan Srivastava}
\email[]{milan.srivastava@sharda.ac.in}
\affiliation{Department of Mathematics, Sharda School of Basic Sciences and Research,\\ Sharda University, Greater Noida, India}
%\date{\today}
%
\begin{abstract}
In the present work, we study a cosmological model composed of a viscous dark matter interacting with decaying vacuum energy in a spatially flat Universe. In the first part, we find the analytical solution of different cosmological parameters by assuming the physically viable forms of bulk viscosity and decaying vacuum density with the interaction term. The second part is dedicated to constrain the free parameters of the interacting viscous model with decaying vacuum energy by employing latest observational data of $Pantheon+$, Cosmic Chronometer and $f(z)\sigma_{8}(z)$. We find that the interacting model just deviate very slightly from well-known concordance $\Lambda$CDM  model and can alleviate effectively the current $H_0$ tension between local measurement by R21 and global measurement by Planck 2018, and the excess in the mass fluctuation amplitude $\sigma_{8}$ essentially vanish in this context. We report the Hubble constants as $H_0=72.150^{+0.989}_{-0.779}$, and $ 72.202^{+0.796}_{-0.937}$ \;$km s^{-1} Mpc^{-1}$, deceleration parameters as $q_0=-0.533 \pm 0.024$, and $-0.531 \pm 0.024$, and equation of state parameters as $w_0=-0.689 \pm 0.016$, and $ -0.687 \pm 0.016$ for $\Lambda$CDM and interacting models, respectively. It is found that the interacting model is in good agreement with $\Lambda$CDM. Further, we discuss the amplitude of matter power spectrum $\sigma_8$ and its associated parameter $S_8$ using $f(z)\sigma_8(z)$ data. Finally, the information selection criterion and Bayesian inference are discussed to distinguish the interacting model with $\Lambda$CDM model.
\end{abstract}

\pacs{98.80.-k, 98.80.Es}
\keywords{Dark energy; Viscous Cosmology; Varying $\Lambda$; Cosmological parameters.}
\maketitle{}

%\tableofcontents
\section{Introduction}
\label{intro}
In modern cosmology, understanding of two dark components: dark energy (DE) and dark matter (DM) of  the Universe is so far the most challenging research area in cosmology. The discovery of accelerated expansion in $1998$ \cite{riess1998,per1999} has motivated to comprehend the composition of these dark entities. Dark matter, which interacts only gravitationally with ordinary matter, explains the measurement of rotation curves of spiral galaxies \cite{persic1996}. The most important theories for DM are scalar fields \cite{magana2012,hernandez2018}, supersymmetry models \cite{martin1998}. On the other hand, dark energy, which has negative pressure, is considered for the accelerated expansion of the Universe. The most interesting candidates are the cosmological constant (CC), phantom energy, quintessence, Chaplygin gas, modified theories of gravity, etc. (refer to Refs.\cite{sahni2000,peebles2003,copeland2006}). The cosmological constant as DE so-called Lambda-cold-dark matter, abbreviated as $\Lambda$CDM is still the best candidate to explain the cosmological observations. However, $\Lambda$CDM model suffers with some theoretical problems \cite{weinberg1989,padmanabhan2003} when its origin is considered as quantum vacuum fluctuations.  \\
\indent Many alternative theories have been proposed beyond the $\Lambda$CDM to solve these conundrums. Some works in literature have shown that some dynamical DE models are able to resolve the problems of $\Lambda$CDM. In this respect, the cosmological models with time-varying vacuum energy are the promising models to resolve the issues. Although there is no fundamental theory to describe a time-varying vacuum, a phenomenological technique has been proposed to parametrized CC. Shapiro and Sol\'{a} \cite{shapiro2002} and Sol\'{a} \cite{sola2008} proposed a running vacuum models (RVMs) on the basis of renormalization group (RG) formalism of Quantum field theory (QFT) in curved spacetime. In the context of RVMs, it is considered that the vacuum energy density evolves slowly with the cosmic expansion. It has been illustrated that both the background and linear perturbation levels of the cosmic evolutions can be described by the RVMs. The vacuum energy is typically determined in curved space-times using renormalization group procedures, which depend on the Hubble parameter $H$ and its time derivative and is given by the form $\rho_{\Lambda} (H, \dot{H}) = M^2_{Pl}\Lambda(H, \dot{H})$ \cite{sola2013}. In past literature, many authors \cite{carvalho1992,lima1994,bertolami1986,ozer1987,peebles1988,overduin1998} have studied the time-varying vacuum energy models. In recent years, the RVMs have been carefully confronted against many cosmological data, which receive a significant success \cite{wang2005,elizalde2005,borges2005,carneiro2006,borges2008,carneiro2008,basilakos2009a,basilakos2009b,costa2010,pigozzo2011,bessada2013,sola2015,sola2017,sola2017a,sola2018a,sola2019a,jayadevan2019,sola2021a,singh2021a,vinita2023}.
\\
\indent On the other hand, viscous fluid cosmology is an interesting research area to understand the accelerated expansion of the Universe. It has been observed that irreversible processes in the evolution of the Universe may also be responsible for explaining the recent accelerated expansion. There are two types of viscosity: bulk and shear, however, bulk viscosity is the one that plays an important role in the evolution of the Universe as it satisfies the cosmological principle. Bulk viscous models have been studied using two approaches: the Eckart \cite{eckart1940} and Israel-Stewart \cite{israel1979} theories. Eckart \cite{eckart1940} proposed a non-causal theory of viscous of first order which was later modified by Landau and Lifshitz \cite{landau1987}. In this theory, all equilibrium states are unstable \cite{hiscock1985} and the signals can propagate through the fluids faster than the speed of light \cite{muller1967}. The Israel-Stewart formalism is a full causal theory of second order which avoids the casuality problem. When the relaxation time goes to zero, the causal theory reduces to the Eckart's first order theory. Despite of the causality problem, Eckart theory is widely used due to its simplicity. In both theories, the bulk viscous term is included in Einstein field equations through an effective pressure, written in the form $\bar{p}=p+\Pi$, where $p$ is pressure of matter contents such as dust-matter, DE or relativistic species (photons and neutrinos) and $\Pi$ refers the bulk viscous pressure. In Eckart theory, the bulk viscous pressure is assumed as $\Pi=-3\zeta H$, where $\zeta$ is the bulk viscous coefficient and $H$ the Hubble parameter. Some works related to Eckart theory have studied the evolution of the Universe at late time by assuming bulk viscous coefficient with a constant \cite{murphy1973,padmanabhan1987,gron1990,maartens1995,brevik2005,norman2017} and polynomial \cite{singh2007,he2009,avelino2010,hernandez2019} as functions of redshift or in terms of energy density. Some more authors \cite{fabris2006,hu2006,ren2006,meng2007,wilson2007,mathews2008,avelino2008,avelino2008a,avelino2009,meng2009,nour2011,
singh2014,mathew2015,bamba2016,wang2017,singh2018,singh2018a,singh2018b,singh2019,singh2020,hu2020} have explored the viability of a bulk viscous Universe to explain the present accelerated expansion of the Universe. \\
\indent Decaying vacuum energy density (VED) models and viscous fluid models are two appealing theoretical models that have been separately studied by many authors to solve some of problems facing by standard $\Lambda$CDM model. Despite the success of decaying vacuum energy and viscous fluid models, it should be noted that they have, separately, limitations in describing the entire cosmological evolution. Recently, in Refs.\cite{herr2020,Ashoorioon2023,cruz2023,singh2024} the authors have studied the cosmological model with combination of these two notions in a single cosmological setting and have investigated their cosmological implications using observational data. \\
\indent The $\Lambda$CDM's problems also motivate research into new physics beyond this standard model. In standard cosmology, it is usually assumed that DM and DE do not interact with each other. However, there is no physical basis for this assumption. In this regard, a popular approach is to investigate the cosmological models where the interactions between the DM and DE take place. In this interaction theory, DM and DE are not separately conserved but they exchange energy (and /or momentum). However, the total energy (and/or momentum) is conserved. It is worth constraining the interacting models against the available wealth of precision cosmological data. This has motivated a large number of studies based on models where DM and DE share interactions, usually referred to as interacting dark energy (IDE) models.  Several studies in the literature have been devoted to explore whether DM-DE interactions may help to resolve the enduring $H_0$ tension \cite{wang2014,nunes2016,valentino2017,valentino2019,wang2018,wang2022,gariazzo2022,westhuizen2024}. In Refs.\cite{chen2011,kremer2012,avelino2012,avelino2013,harko2013,leyva2017,atreya2018,hernandez2020}, authors have studied the interacting viscous dark fluids for explaining the early and late time evolution of the Universe. The above cited works on bulk viscosity show that interacting viscous fluid models can be one of the possible mechanism to explain the present acceleration of the Universe.\\
\indent In light of the aforementioned discussions, the main purpose of the present work is to investigate a cosmological model for a spatially flat Friedmann-Lema\^{i}tre-Robertson-Walker Universe including two components: a non-perfect and interacting viscous dark matter, and decaying vacuum energy that interact with viscous dark matter in Eckart's approach. We analyze the dynamics of interacting model by constraining the free parameters by performing a Markov chain Monte Carlo (MCMC) using the latest observational data. Further, we investigate how our interacting viscous model with decaying VED affects the perturbation level. To investigate the role of this model in structure formation, we employ the perturbation equation to determine the growth of matter fluctuations. Finally, we study the evolutions of various cosmological parameters and compare the perfect fluid case that corresponds to the $\Lambda$CDM model through the model selection criteria and Bayesian evidence analysis.\\
\indent The paper is organised as follows. In Sec. II, we present the general features of the proposed cosmological model for a spatially flat Friedmann-Lema\^{i}tre-Robertson-Walker Universe where dissipative effects are present with interacting decaying VED. Section III deals with the analysis on Structure formation and perturbation equations. We present in Sec. IV the cosmological probes that are used to constrain the model. Section V gives the results and discussions on various cosmological parameters with the trajectories and compares the proposed model with the concordance $\Lambda$CDM using information criterion and the Bayesian inference. The main results of the work are summarized and discussed in Sec. VI. Two more solutions are also presented in Appendix.
\section{Interacting bulk viscous model with decaying VED}
\label{sec:1}
We start with a homogeneous and isotropic Universe described by the Friedmann-Lema\^{i}tre-Robertson-Walker (FLRW) metric
\begin{equation}\label{eq1}
ds^2 = -dt^2 + a^2(t) \left[dr^2+r^2(d\theta^2+sin^2\theta d\phi^2)\right],
\end{equation}
where $a(t)$ represents the cosmic scale factor of the Universe. The Einstein field equations with a time-varying cosmological constant term, $\Lambda$ take the form
\begin{equation}\label{eq2}
R_{\mu\nu}-\frac{1}{2}g_{\mu\nu}R =8\pi G(T_{\mu\nu}+g_{\mu\nu}\rho_{\Lambda})
\end{equation}
where $R = g_{\mu\nu}R_{\mu\nu}$ is the Ricci scalar, $\rho_{\Lambda}=\Lambda/8\pi G$ is the energy density associated to $\Lambda$-term, so-called the vacuum energy density (VED), $G$ is the Newton gravitational constant and $T_{\mu\nu}$ is the energy-momentum tensor of matter. It should be emphasised that we employ the geometric units $8\pi G = c =1$. \\
\indent In this paper, we propose to study the cosmological dynamics of the Universe which include the interaction between the dark matter component including dissipation through a bulk viscous coefficient and a vacuum energy density described by running coupling depending on the Hubble parameter (hereafter we refer interacting viscous $\Lambda(t)$ model). Then, the energy-momentum tensor is given by
\begin{equation}\label{eq3}
T_{\mu\nu}=(\rho+\tilde{p})u_{\mu}u_{\nu}+g_{\mu\nu}\tilde{p}
\end{equation}
where  $\rho=\rho_m+\rho_{\Lambda}$ is the sum of total energy density of the fluid contributed from $\rho_m$, the energy density of DM and $\rho_{\Lambda}$, the energy density of vacuum, $u_{\mu}$ is the associated four-velocity and $\tilde{p}=p_m+\Pi+p_{\Lambda}$ is the sum of total pressure of fluid contributed from $p_m$, the equilibrium pressure, $\Pi$, the non-equilibrium pressure due to bulk viscosity and $p_{\Lambda}$, the pressure due to vacuum. It is assumed that deviation of any system from the local thermodynamical equilibrium is the source of bulk viscosity. In accordance with the second law of thermodynamics, the restoration of thermal equilibrium is a dissipative process that produces entropy. As a result of this entropy generation, the bulk viscous term causes an expansion in the system.\\
\indent The viscous fluid in homogeneous and isotropic cosmological models is determined by its bulk viscosity. The Eckart's formalism serves as the basis for this theory\cite{eckart1940}. It is basically obtained from the second order theory of non-equilibrium thermodynamics in the limit of vanishing relaxation time which was proposed by Israel and Stewart \cite{israel1979}. Inspired by the viscosity behavior in fluid mechanics, being proportional to the speed, we assume $\Pi = -3\zeta H$, where $H$ is the Hubble parameter and $\zeta$ is the bulk viscous coefficient, which is assumed to be positive on thermodynamic grounds. Furthermore, we consider the non-relativistic matter with $p_m=0$ to be the bulk viscous fluid. Therefore, the sum of the vacuum energy pressure, $p_{\Lambda}=-\rho_{\Lambda}$ and viscous pressure $\Pi=-3\zeta H$ are the components contributing to the total pressure. These two extra ingredients have been introduced to get a more realistic fluid description of DM and DE and also a suitable comparison with $\Lambda$CDM model.\\
\indent In the presence of a non-gravitational interaction between viscous dark matter and decaying vacuum energy, which is characterized by a coupling function $Q(t)$, also known as the interacting rate, the  friedmann equation and conservation equations can be written as
\begin{equation}\label{eq4}
3H^2 =  \rho_{m} + \rho_{\Lambda},
\end{equation}
\begin{equation}\label{eq5}
\dot{\rho}_{m}+3H\rho_{m} = 9\zeta H^2+Q(t),
\end{equation}
\begin{equation}\label{eq6}
\dot{\rho}_{\Lambda}+3H(\rho_{\Lambda}+p_{\Lambda})= -Q(t),
\end{equation}
\noindent where dot denotes derivative with respect to the cosmic time $t$. In Eqs.\eqref{eq5} and \eqref{eq6}, $Q (t)$ denotes the interaction function providing the rate of energy transfer between viscous dark matter and decaying VED. It is to be noted that $Q(t)<0$ gives energy transfer from viscous DM to decaying vacuum energy where as $Q(t)>0$ gives energy transfer from decaying vacuum energy to viscous DM. Once the interaction function $Q(t)$ is specified, the background dynamics of the model can be found using \eqref{eq4}-\eqref{eq6}. We can define $Q(t)$ by two ways: either by deriving the interaction function from some fundamental physics at the Lagrangian level or by assuming at phenomenological level and testing using observational data. Due to the absence of a fundamental physical theory, we will consider the second approach by assuming the interaction function as \cite{nojiri2011,brevik2015}
\begin{equation}\label{eq7}
Q = 3\alpha \rho_{m}H,
\end{equation}
where the term $\alpha$ denotes the dimensionless coupling parameter and included in the fitting vector of free parameters which is to be constrained by the observational dataset.\\
\indent For decaying VED, we assume a phenomenological application of renormalization group analysis, which can be written as \cite{sola2017,vinita2023}
\begin{equation}\label{eq8}
\rho_{\Lambda}=c_0+3\nu H^2,
\end{equation}
where $c_0=3H^2_0(\Omega_{\Lambda}-\nu)$ is the additive constant and fixed by the boundary condition $\rho_{\Lambda}(H_0)=\rho_{{\Lambda}0}$. In this scenario, $\nu$ is the dimensionless vacuum parameter and is naturally anticipated to have extremely small magnitude i.e. $|\nu|\ll1$. Thus, the positive magnitude of $\nu$ enables the vacuum's cosmic evolution. In this instance, we will fit $\nu$ to the cosmological data set by taking it as a free parameter.\\
\indent Substituting \eqref{eq7} into \eqref{eq5}, we get
\begin{equation}\label{eq9}
\dot{\rho}_{m} = 9 \zeta H^2 -3(1-\alpha)\rho_{m}H
\end{equation}
From \eqref{eq4}-\eqref{eq9}, we get the following Hubble evolution equation,
\begin{equation}\label{eqh}
\dot{H}+\frac{3}{2}(1-\alpha)H^2 =  \frac{3}{2} \frac{\zeta}{(1-\nu)} H + \frac{1}{2}\left(\frac{1-\alpha}{1- \nu}\right)c_0.
\end{equation}
The above evolution equation \eqref{eqh} has $H$ and $\zeta$ as unknown quantities. We can get the solution of $H$ only if the functional form of $\zeta$ is specified. We consider the bulk viscous coefficient $\zeta$ as proportional to the expansion rate of the Universe, i.e., to the Hubble parameter $H$, which can be expressed as \cite{gron1990,meng2007,singh2007,singh2014,brevik2002a,huan2020}
\begin{equation}\label{eq12}
\zeta=\zeta_1 H
\end{equation}
where $\zeta_1$ is a dimensionless constant to be estimated from the observations. Substituting this relation into \eqref{eqh}, we get
\begin{equation}\label{eqb1}
\dot{H}+\frac{3}{2}\left(\frac{(1-\nu)(1-\alpha)-\zeta_1}{1-\nu}\right)H^2- \frac{1}{2}\left(\frac{1-\alpha}{1- \nu}\right)c_0 = 0.
\end{equation}
The preceding equation with a variable change from $t$ to $x=\log a$ together with $c_0=3H^2_0(\Omega_{\Lambda }-\nu)$ can be written as
\begin{equation}\label{eqb2}
\frac{dh^2}{dx}+3\left(\frac{(1-\nu)(1-\alpha)-\zeta_1}{1-\nu}\right)h^2=3\left(\frac{(\Omega_{\Lambda}-\nu)(1-\alpha)}{1-\nu} \right),
\end{equation}
where $h=H/H_0$ represents the Hubble parameter which is dimensionless and $\Omega_{\Lambda}=\rho_{\Lambda}/3H^2_0$. Assuming $(1-\nu)(1-\alpha)-\zeta_1>0$ and employing the redshift relation to the normalised scale factor, $a=(1+z)^{-1}$,  Eq.\eqref{eqb2} gives
\begin{equation}\label{eqb3}
E^2(z)=\tilde{\Omega}_{m,0}(1+z)^{3\left(\frac{(1-\nu)(1-\alpha)-\zeta_1}{(1-\nu)}\right)}+\tilde{\Omega}_{{\Lambda},0},
\end{equation}
where
\begin{equation}
\tilde{\Omega}_{{\Lambda},0}= \frac{(\Omega_{\Lambda}-\nu)(1-\alpha)}{(1-\nu)(1-\alpha)-\zeta_{1}},
\end{equation}
and
\begin{equation}
\tilde{\Omega}_{m,0}=1-\frac{(\Omega_{\Lambda}-\nu)(1-\alpha)}{(1-\nu)(1-\alpha)-\zeta_{1}}=1-\tilde{\Omega}_{{\Lambda},0}.
\end{equation}
\noindent In Eq.\eqref{eqb3}, $E(z)=H/H_0$ is dimensionless Hubble parameter and $\Omega_{i,0}$ ($i=\text{DM,DE}$) represents the current value of density parameter of viscous DM and decaying vacuum energy. For $E(0)=1$, we have $\tilde{\Omega}_{m,0}+\tilde{\Omega}_{{\Lambda},0}=1$. The scale factor of expansion can be calculated by using  $H=\dot{a}/a$ and integrating the equation \eqref{eqb3}, the solution for the scale factor $a(t)$ is given by
\begin{equation}\label{at}
a(t)=\left(\frac{\tilde{\Omega}_{m,0}}{\tilde{\Omega}_{{\Lambda},0}}\right)^{\frac{(1-\nu)}{3\left[(1-\alpha)(1-\nu)-\zeta_1\right]}}
\left(\sinh^{2/3}\left(\frac{t}{\tilde{t}}\right)\right)^{\frac{(1-\nu)}{(1-\alpha)(1-\nu)-\zeta_1}},
\end{equation}
where $\tilde{t}\equiv 2/(3H_0\sqrt{{\frac{\left[(1-\nu)(1-\alpha)-\zeta_{1}\right](1-\alpha)(\Omega_{\Lambda}-\nu)}{(1-\nu)^2}}})$. It can be observed from \eqref{at} that the scale factor reduces to $a(t)=(\Omega_{m,0}/\Omega_{{\Lambda},0})^{1/3}\;\sinh^{2/3}(t/\tilde{t})$ for $\alpha=0$, $\zeta_1=0$ and $\nu=0$, which is the analytical solution of the scale factor for $\Lambda$CDM model. It can be observed from above equation that the scale factor varies as $a\propto t^{\frac{2(1-\nu)}{3[(1-\nu)(1-\alpha)-\zeta_1]}}$ during early times which the power-law expansion of the Universe. For late-time evolution the scale factor varies as $a\propto \exp{\sqrt{\frac{(\Omega_{\Lambda}-\nu)}{3(1-\zeta_1-\nu)}}H_0 t}$, which implies the de Sitter Universe.\\
\indent To investigate the decelerated and accelerated phases of the expansion of the Universe as well as its transition during the evolution, we explore a crucial cosmological parameter, called `deceleration parameter'. The definition of deceleration parameter $q$ is
\begin{equation}\label{eqq1}
q=-\frac{\ddot{a}}{a}\frac{1}{H^2}=-\left(1+\frac{\dot{H}}{ H^2}\right).
\end{equation}
\indent Using \eqref{eqb3} into \eqref{eqq1}, the deceleration parameter $q$ in terms of redshift $z$ is given by
\begin{widetext}
\begin{equation}\label{eqb7}
q(z) = -1 + \frac{3}{2(1-\nu)}\left(\frac{((1-\alpha)(1-\Omega_{\Lambda})-\zeta_{1})(1+z)^{3\left(\frac{(1-\nu)(1-\alpha)-\zeta_1}{(1-\nu)}\right)}}{\frac{(\Omega_{\Lambda}-\nu)(1-\alpha)}{(1-\nu)(1-\alpha)-\zeta_{1}}
 + \left(1- \frac{(\Omega_{\Lambda}-\nu)(1-\alpha)}{(1-\nu)(1-\alpha)-\zeta_1}\right)(1+z)^{3\left(\frac{(1-\nu)(1-\alpha)-\zeta_1}{(1-\nu)}\right)}}\right)
\end{equation}
\end{widetext}
The aforementioned equation demonstrates how the redshift affects the dynamics of $q$ . We note that the value of $q(z)$ approaches $-1$ in the future (negative redshift).
Furthermore, we determine the present value of $q$ for $z=0$, denoted as $(q_{0})$, which is given by
\begin{equation}\label{eqb8}
q_0=-1+\frac{3}{2(1-\nu)}\left[(1-\alpha)(1-\Omega_{\Lambda})-\zeta_{1}\right].
\end{equation}
The accelerating and decelerating phase of the Universe can be inferred from the negative and positive signs of the deceleration parameter, respectively. For $q_{0}> 0$, the Universe exhibits expanding behaviour and goes through a deceleration phase. For $-1<q_{0}<0$, represents the current state of the Universe which is the expanding and accelerating Universe.\\
\indent For the sake of completion, we further calculate effective equation of state parameter $w_{eff}$ as a function of redshift $z$,
\begin{widetext}
\begin{equation}\label{eqb8}
w_{eff}(z)= -1 + \frac{1}{(1-\nu)} \left(\frac{((1-\alpha)(1-\Omega_{\Lambda})-\zeta_{1})(1+z)^{3\left(\frac{(1-\nu)(1-\alpha)-\zeta_1}{(1-\nu)}\right)}}{\frac{(\Omega_{\Lambda}-\nu)(1-\alpha)}{(1-\nu)(1-\alpha)-\zeta_{1}}
 + \left(1- \frac{(\Omega_{\Lambda}-\nu)(1-\alpha)}{(1-\nu)(1-\alpha)-\zeta_1}\right)(1+z)^{3\left(\frac{(1-\nu)(1-\alpha)-\zeta_1}{(1-\nu)}\right)}}\right)
\end{equation}
\end{widetext}
\indent At $z=0$, the present value of $w_{eff}$ is determined by
\begin{equation}\label{eqb11}
w_{eff}(z=0)=-1+\frac{1}{(1-\nu)}\left[(1-\alpha)(1-\Omega_{\Lambda})-\zeta_{1}\right].
\end{equation}
\section{Growth of perturbations in interacting model}\label{perturbation}
The presence of cosmological fluctuations influences the background cosmology in which the perturbations evolve. Hence for the complete analysis of our interacting viscous $\Lambda(t)$ model we must take into account the effects on the large scale structure $(LSS)$ formation data, which we incorporate into our observational analysis, in order to fully confront the model. As a result, we must take the matter density perturbations into consideration. Details on this portion of the analysis has been provided in many references \cite{sola2018,agvalent2018}. Here, we simply cite the resulting differential equation, which is entirely consistent with the analysis of \cite{singh2024} and references therein.  For the interacting viscous $\Lambda(t)$ model, we use the standard perturbations equation for the linear matter density contrast $\delta_m\equiv \delta\rho_m/\rho_m$ as given below:
\begin{equation}\label{p}
\delta_{m}''+ \left( \frac{3}{a} + \frac{H'(a)}{H(a)} \right)\delta_{m}' - \frac{4\pi G \rho_m}{H^2(a)} \frac{\delta_{m}}{a^2} = 0
\end{equation}
Here $()'\equiv d/da$ represents the derivative with respect to the scale factor. The aforementioned approximation to structure formation is sufficient for considering the primary consequences originate from the distinct expression of the Hubble function as compared to the $\Lambda$CDM. Thus, the preceding second-order differential equation \eqref{p} seems to be accurate. We examine the Hubble function \eqref{eqb3} as it was determined in Section II. The smoothness of the matter perturbation in the interacting viscous $\Lambda(t)$ model is described by the equation \eqref{p}.\\
\indent The weighted linear growth, $f(z)\sigma_8(z)$, is typically used to compare the theoretical calculations with the structure formation data in the linear regime. Here, $\sigma_8(z)$ represents the r.m.s. mass fluctuations on $R_8 = 8 h^{-1}$ Mpc at redshift $z$ and $f(z)= d \ln \delta_{m}(a)/ d\ln(a)$ is the growth factor. The details of the calculation are provided in coming Section-IV(C).
\section{Data and methodology}
In this section, we use a variety of observational data and methodology in order to constrain the model parameters of $\Lambda$CDM and interacting viscous $\Lambda$(t) models which comprise observations from: (i) Pantheon$+$ dataset; (ii) Hubble dataset(Cosmic Chronometers); and (iii) $f(z)\sigma_{8}(z)$ dataset. The following subsections provide an overview of each data set.
\subsection{Pantheon$+$ dataset}
We consider an updated Pantheon+ sample \cite{scolnic2022} which is the successor of the Pantheon sample. This compilation comprises of $1701$ light curves gathered by $1550$ type Ia supernovae (SNe), providing detailed information for cosmological analysis. In addition to the most recent SNe sightings, the Pantheon+ collection expands upon earlier SNe compilations and offers a broad range of redshifts ranging from $z= 0.00122$ to $2.2613$.  The Chi-squared related to the Pantheon$+$ dataset is given by
\begin{equation}\label{pan}
\chi^2_{Pan+} =  \sum_{i=1}^{1701} \vec{D} ^{T} C^{-1}_{Pan+} \vec{D}
\end{equation}
where $  \vec{D}= m_{Bi} - M - \mu_{model}$. Here, $m_{Bi}$ and $M$ are the observed peak magnitude in the rest frame of the $B$ band and absolute $B$-band magnitude respectively. The theoretical distance modulus $\mu_{model}$ is defined by
\begin{equation}
 \mu_{model}(z_{i}) = 5 \log_{10}\left(\frac{D_{L}(z_{i})}{1 M pc} \right) + 25,
 \end{equation}
where $D_{L}(z)$ represents the luminosity distance, that is described by $D_{L}(z) = (1 + z) H_0 \int_0^{z} \frac{ dz'}{H(z')}$.
In Eq. \eqref{pan}, $C$ denotes the total covariance matrix which takes the form $C=D_{stat}+C_{sys}$, where the diagonal matrix $D_{stat}$ and covariant matrix $C_{sys}$ denote the statistical uncertainties and the systematic uncertainties.\\
\indent In contrast to the Pantheon dataset, the degeneracy between the absolute magnitude $M$ and $H_0$ is broken in $Pantheon+$. In order to accomplish this, the vector $\vec{D}$ in Eq. \eqref{pan} is written in terms of the distance moduli of SNe in Cepheid hosts. As a result, $M$ can have an independent constraint, which leads to the following expression:
\begin{equation}
\vec{D'} =
\begin{cases}
  m_{Bi} - M - \mu_{i}^{Cepheid}&  i \in Cepheid  \\
  m_{Bi} - M - \mu_{model}(z_i) & Otherwise \\
\end{cases}
\end{equation}
where $\mu_{i}^{Cepheid}$ denotes the distance modulus, determined independently using Cepheid calibrators, and corresponds to the Cepheid host of the $i^{th}$ SNe.  Hence, we can rewrite Eq.\eqref{pan} as follows:
\begin{equation}\label{pn}
\chi^2_{Pan+} =  \sum_{i=1}^{1701} \vec{D'} ^{T} C^{-1}_{Pan+} \vec{D'}
\end{equation}
\subsection{Cosmic chronometers}
The cosmic chronometers (CC) is another observational data obtained through the differential-age method. We can find the Hubble parameter values from CC method at distinct redshifts based on the relative age of passively evolving galaxies. For our analysis, we employ a compilation of $32$ data points of the Hubble parameter derived through the differential age technique \cite{moresco2022} in the redshift range $0.07 \leq z \leq 1.965 $ with errors. The Chi-squared function for CC is given by
\begin{equation}
\chi_{CC}^2=\sum_{i=1}^{32}\frac{[H(z,\theta)-H^{obs}(z_{i})]^2}{\sigma^2_{H(z)}}
\end{equation}
where $H(z)$ represents the theoretical prediction of the Hubble parameter calculated in Eq. \eqref{eqb3} and $H^{obs}(z)$ are the observational data with errors $\sigma_{H(z)}$.
\subsection{ $f(z)\sigma_8(z)$ data}
Finally, in this work we use the recent ``Gold -17" compilation consisting of $18$ independent measurements of $f(z)\sigma_8(z)$. These data points are based on Redshift Space Distortion (RSD) measurements from various observations of the Large Scale Structure (LSS) and complied in Table III of Ref. \cite{nesseris2017}. \\
\indent The linear growth rate of matter perturbations is defined as follows:
\begin{equation}\label{f1}
f(z)=\frac{d\ln \delta_m}{d \ln a} \equiv -(1+z)\frac{\delta_m'(z)}{\delta_m(z)}
\end{equation}
Here, (') denotes the derivative with respect to redshift. Further, in the linear regime $\sigma_8(z)$ is the redshift-dependent parameter which measures the growth of root-mean-square mass fluctuations in spheres with radius $8h^{-1}$ Mpc scales which is given as:
\begin{equation}\label{f2}
\sigma_8(z)=\sigma_8(z=0)\frac{\delta_m(z)}{\delta_m(z=0)}
\end{equation}
Combining \eqref{f1} and \eqref{f2}, the weighted linear growth is determined by:
\begin{equation}\label{f3}
f\sigma_8(z)= -(1+z)\frac{\sigma_8(z=0)}{\delta_m(z=0)} \frac{d\delta_m}{dz}
\end{equation}
The Chi-squared function of $f(z)\sigma_8(z)$ can be calculated as \cite{quelle2020}:
\begin{equation}\label{f4}
\chi_{f(z)\sigma_8(z)}^2=\sum_{i=1}^{18}\frac{[f\sigma_8(z_i)-(f\sigma_{8,i})_{ob}]^2}{\sigma_{i}^2}
\end{equation}
\indent Based on the EMCEE \cite{foreman2013} python module, we employ the Markov Chain Monte Carlo(MCMC) statistical technique to examine the parameter space of our cosmological models and minimise the $\chi^2$ function for both $\Lambda$CDM and interacting viscous $\Lambda(t)$ models. We consider the joint analysis by assuming the sum of all $\chi^2$ functions:
\begin{equation}
\chi^2_{tot}= \chi^2_{Pan+}+\chi^2_{CC}+ \chi^2_{f\sigma_8(z)}.
\end{equation}
To perform this analysis, we choose uniform priors for the parameters of models which is listed in Table I.
%\begin{widetext}
\begin{table}
\caption{Flat prior used on various parameters during statistical analysis.}
\begin{tabular}{|c||c|}
\hline
Parameters & Priors\\
\hline
$H_0$ & $[50, 100]$\\

$\Omega_{\Lambda}$ & $(0, 1]$\\
 $\zeta_1$ & $(0, 1]$ \\
 $\nu$ & $(0, 1]$\\
  $\alpha$ & $(0, 1]$\\
   $\sigma_8$ & $[0.6, 1.2]$\\
\hline

\end{tabular}
\label{table:values1}
\end{table}
%\end{widetext}
\begin{table}
\caption{The values of parameters for $\Lambda$CDM and interacting viscous $\Lambda(t)$ models for combination of Pantheon$+$, Hubble(CC) and $f(z)\sigma_{8}(z)$ observational datasets.}

\begin{tabular}{ |l||c|c|}

%\multicolumn{4}{|c|}{Viscous $\Lambda$(t)} \\
\hline
Parameter & $\Lambda$CDM & Interacting viscous $\Lambda(t)$  \\
\hline
$H_0$  & $72.150^{+0.989}_{-0.779}$ & $72.202^{+0.796}_{-0.937}$   \\

$\Omega_{\Lambda}$ & $0.689^{+0.016}_{-0.017}$ & $0.681^{+0.014}_{-0.017}$  \\

$M$ & $-19.292^{+0.026}_{-0.023}$ & $-19.290^{+0.024}_{-0.026}$  \\

$\zeta_1$ & $-$ & $0.010^{+0.006}_{-0.007}$   \\

$\nu$ & $-$ & $0.005^{+0.003}_{-0.003}$ \\	

$\alpha$ & $-$ & $-0.005^{+0.003}_{-0.004}$   \\

$\sigma_8$ & $0.750^{+0.034}_{-0.031}$  &  $0.739^{+0.034}_{-0.030}$  \\

$S_8$ & $0.765^{+0.030}_{-0.030}$  &  $0.762^{+0.030}_{-0.030}$  \\

$q_{0}$  & $-0.533^{+0.024}_{-0.024}$ & $-0.531^{+0.024}_{-0.024}$  \\

$z_{tr}$  & $0.643^{+0.041}_{-0.041}$ & $0.647^{+0.038}_{-0.044}$ \\

$w_{0}$  & $-0.689^{+0.016}_{-0.016}$ & $-0.687^{+0.016}_{-0.016}$ \\

$t_0(Gyr)$ & $13.74^{+0.019}_{-0.021}$ & $13.75^{+0.023}_{-0.021}$  \\

\hline
\end{tabular}
\label{table:values1}
\end{table}

\begin{figure*}
\begin{minipage}[t]{0.32\linewidth}
\includegraphics[width=\linewidth]{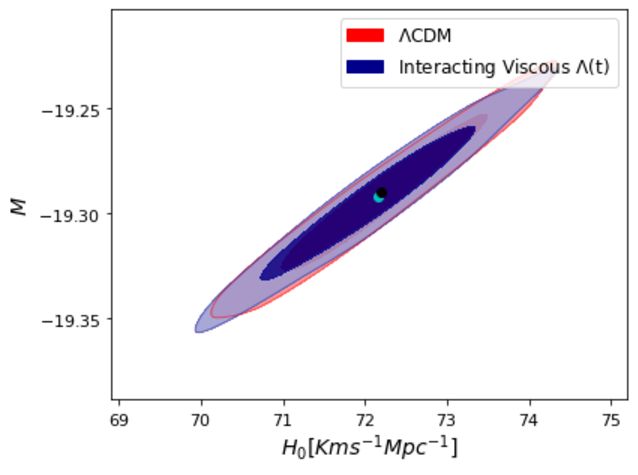}
\end{minipage}
\begin{minipage}[t]{0.32\linewidth}
\includegraphics[width=\linewidth]{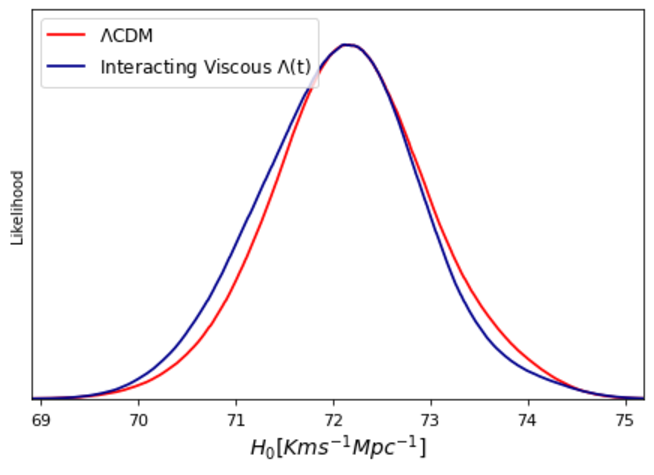}
\end{minipage}
\begin{minipage}[t]{0.32\linewidth}
\includegraphics[width=\linewidth]{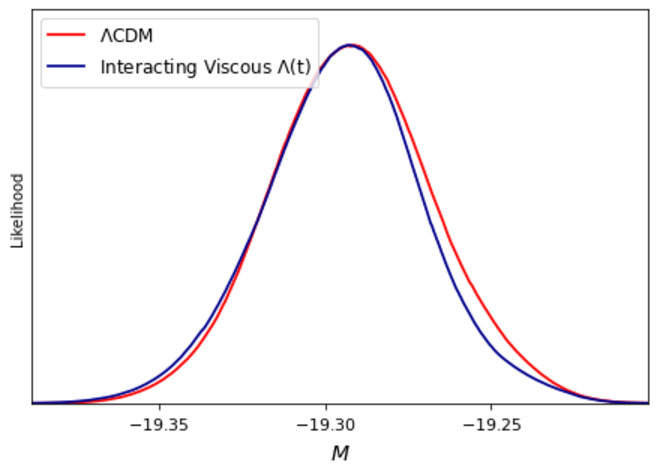}
\end{minipage}
\caption{The $68.3\%$ (dark shaded) and $95.4\%$ (light shaded) confidence contours and likelihoods for $H_0$ and  $M$ for the $\Lambda$CDM and interacting viscous $\Lambda(t)$ models using combined dataset. The best-fit values for these models are represented by the cyan and black dot respectively.}
\end{figure*}\label{fig:4}

\begin{figure*}
\begin{minipage}[t]{0.32\linewidth}
\includegraphics[width=\linewidth]{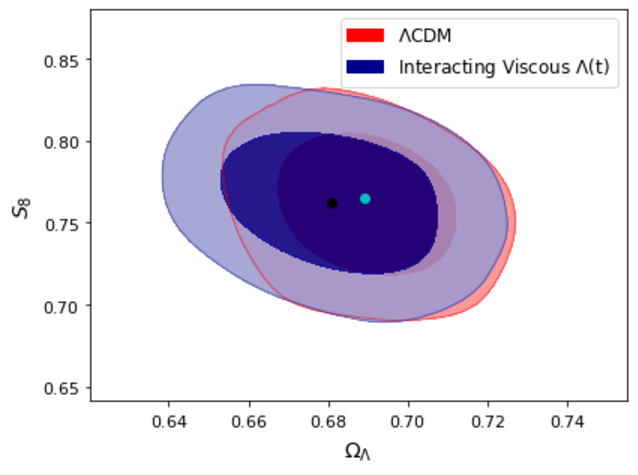}
\end{minipage}
\begin{minipage}[t]{0.32\linewidth}
\includegraphics[width=\linewidth]{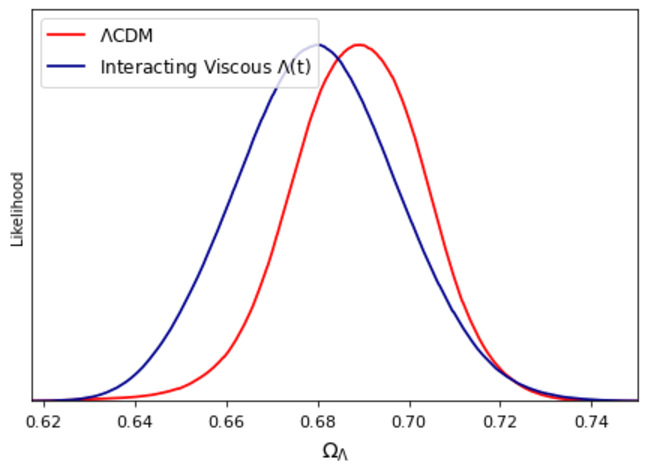}
\end{minipage}
\begin{minipage}[t]{0.32\linewidth}
\includegraphics[width=\linewidth]{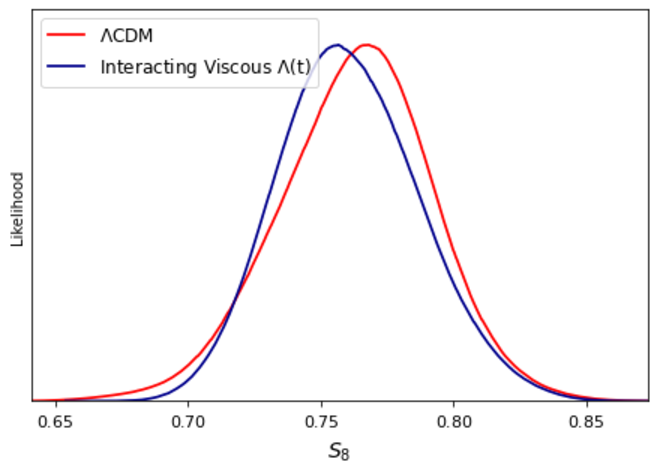}
\end{minipage}
\caption{The $68.3\%$ (dark shaded) and $95.4\%$ (light shaded) confidence contours and likelihoods for $\Omega_{\Lambda}$ and  $S_{8}$ for the $\Lambda$CDM and interacting viscous $\Lambda(t)$ models using combined dataset. The best-fit values for these models are represented by the cyan and black dot respectively.}
\end{figure*}\label{fig:5}
\begin{figure*}
\begin{minipage}[t]{0.32\linewidth}
\includegraphics[width=\linewidth]{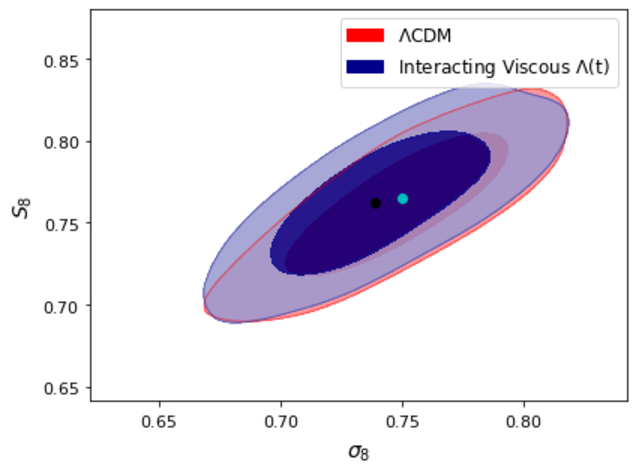}
\end{minipage}
\begin{minipage}[t]{0.32\linewidth}
\includegraphics[width=\linewidth]{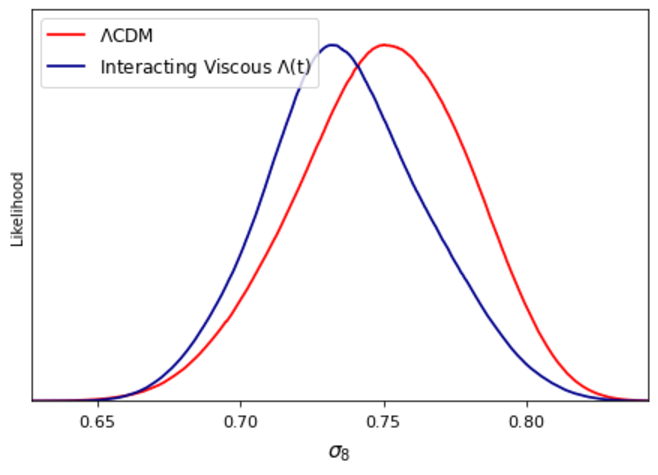}
\end{minipage}
\caption{The $68.3\%$ (dark shaded) and $95.4\%$ (light shaded) confidence contours and likelihoods for $S_{8}$ and  $\sigma_{8}$ for the $\Lambda$CDM and interacting viscous $\Lambda(t)$ models using combined dataset. The best-fit values for these models are represented by the cyan and black dot respectively.}
\end{figure*}\label{fig:4}
\section{Result and discussion}\label{result}
In this section, we present the results of our observational analysis on $\Lambda$CDM and interacting viscous $\Lambda(t)$ models, including both the constraints and cosmological parameters from the current data. In the first subsection, we will present the parameters constraints achieved by observational analysis using the data discussed in Sect.IV. In second subsection, we will implement the Bayesian inference, and in last subsection, we will examine the model comparison through information criterion.\\
\subsection{Parameter constraints}
The best-fit values of parameters of $\Lambda$CDM and interacting viscous $\Lambda(t)$ models using a combined $CC+f(z)\sigma_8(z)+Pan+$ data are summarized in Table II. Figures 1-3 show the $68.3\%$ and $95.4\%$ confidence regions and marginalized likelihood distributions for $\Lambda$CDM and interacting viscous $\Lambda(t)$ models, respectively. The GetDist code \cite{lewis2019} is utilised to retrieve their mean values and the aforementioned Figs. 1-3. In what follows we discuss the constraints on different cosmological as well as model parameters.\\
\indent It is noted that the $H_0$ determined by Riess et al.\cite{riess2022} is $73.04 \pm 1.04$ $km s^{-1} Mpc^{-1}$, so-called R21 where as Planck Collaboration \cite{plank2020} predicts $H_0=67.4 \pm 0.5$  $km s^{-1} Mpc^{-1}$ at $5\sigma$ confidence level. This discrepancy is recognized as ``Hubble tension". For $\Lambda$CDM  model we obtain $H_0= 72.150^{+0.989}_{-0.779}$ $km s^{-1} Mpc^{-1}$. This is in tension with both Planck \cite{plank2020} and R21 \cite{riess2022} results at $4.67\sigma$ and $0.65\sigma$ respectively. For interacting viscous $\Lambda(t)$ model, we obtain $H_0 = 72.202^{+0.796}_{-0.937}$ $km s^{-1} Mpc^{-1}$ which is tension with both Planck and R21 results at $4.80\sigma$ and $0.86\sigma$ respectively. In other words, the $H_0$ measurement of interacting viscous $\Lambda(t)$ model is consistent with R21. The evolutions of $H(z)$ with redshift $z$ are shown in Fig.4 which predict that the trajectories cover majority of dataset with error bars of $H(z)$. This means that the interacting viscous $\Lambda(t)$ model agrees with the combination of dataset as well as $\Lambda$CDM model.\\
\indent In the interacting viscous $\Lambda(t)$ model, we obtain the viscosity coefficient, $\zeta_1$ as $0.010^{+0.006}_{-0.007}$ and $\nu$ as $0.005^{+0.003}_{-0.003}$. The coupling parameter of the interaction term, $\alpha$ is calculated as $-0.005^{+0.003}_{-0.004}$. The negative value of $\alpha$ indicates that there is a possibility of the dark matter to decay into the dark energy.\\
\indent Using the best-fit values, the evolutions of deceleration parameter with redshift for $\Lambda$CDM and interacting viscous $\Lambda(t)$ models with errors are plotted in Fig.5. The trajectories show that the models have transition from decelerating phase to accelerating phase at transition redshift $z_{tr}=0.643^{+0.041}_{-0.041}$ and $z_{tr}=0.647^{+0.038}_{-0.044}$, respectively. The present value of deceleration parameter are found to be $q_0=-0.533^{+0.024}_{-0.024}$ and $q_{0} = -0.531^{+0.024}_{-0.024}$. These values of $q_0$ are within the range of the observational results, i.e., $q_0=-0.64\pm 0.12$ \cite{plank2020}. In late time of evolution, $q(z)$ approaches to  $-1$ in both models which is the future de Sitter phase.\\
\indent In Fig.6, we plot the effective EoS parameter as a function of redshift for best-fit values of $\Lambda$CDM and interacting viscous $\Lambda(t)$ models. It is observed that  $w_\text{eff} \rightarrow -1$ in the late-time in both the models, which implies that the models correspond to de Sitter phase in late-time. Using the best-fit, we get the present EoS parameters $w_\text{eff}(z=0)=-0.689\pm 0.016$ and $w_\text{eff}(z=0)=-0.687\pm 0.016$, respectively.\\
\indent Next, we explore another tension between the theoretical prediction of the growth rate of matter perturbations with the observational growth rate data points for $\Lambda$CDM and interacting viscous $\Lambda(t)$ models. For both models the constraints on $\sigma_8$ and $S_8$ are given in Table II. The amplitude of the matter power spectrum ($\sigma_8$) and its associated parameter $S_8 = \sigma_8 \sqrt{(1-\Omega_\Lambda)/0.3}$ may be determined using the $f(z)\sigma_8(z)$ dataset and it is further possible to calculate the $\sigma_8/ S_8$ tension. The combined dataset gives $\sigma_{8} = 0.750^{+0.034}_{-0.031} $ and $S_{8} = 0.765^{+0.030}_{-0.030}$ in the $\Lambda$CDM model and $\sigma_8= 0.739^{+0.034}_{-0.030}$ and $S_8 = 0.762^{+0.030}_{-0.030} $ in interacting viscous $\Lambda(t)$ model. Our results are perfectly consistent with the combined use of the SDSS and KiDS/Viking data which gives $\sigma_{8} = 0.760^{+0.025}_{-0.020}$ and $S_{8} = 0.766^{+0.020}_{-0.014}$ \cite{heymans2021}. The interacting viscous $\Lambda(t)$ model is in $2.21\sigma$ tension in $\sigma_8$ and $2.11\sigma$ tension in $S_8$ with the corresponding values of Plank \cite{plank2020}. These tensions are not as large compared to $H_0$ tension as discussed above. The confidence contours in ($S_8, \Omega_{\Lambda}$) and ($S_8, \sigma_8$) parameter space and corresponding likelihoods are shown in Figs. 2 and 3, which show no deviation from $\Lambda$CDM contours. This confirms that the measurement of $\sigma_8$ and $S_8$ using $CC+Pan+$ with $f(z)\sigma_8(z)$ data is fully consistent with $\Lambda$CDM. The trajectories of $f(z)\sigma_8(z)$ for $\Lambda$CDM and interacting viscous $\Lambda(t)$ models are plotted in Fig.7. It can be observed that both the models are consistent with the observational data points.\\
\indent Let us discuss the interacting viscous $\Lambda(t)$ model by defining a dimensionless parameter, known as jerk parameter, $j$. This parameter is purely kinematical and is associated with the third order derivative of the scale factor $a(t)$. It is defined as $j =\frac{\dddot{a(t)}}{aH^3}$ \cite{mamon2018}. In terms of the deceleration parameter, it can be expressed as follows \cite{mamon2018}:
\begin{equation}
j(z)=q(2q+1)+(1+z)\frac{dq}{dz}.
\end{equation}
It can provide us the deviation of any model from the $\Lambda$CDM model. It is noted that jerk parameter always has constant value $j=1$ for $\Lambda$CDM model. We report $j_0=0.992$ in interacting viscous $\Lambda(t)$ model which is quite comparable to the $\Lambda$CDM model. We plot the evolution of jerk parameter with respect to the redshift as shown in Fig.8. We observe that the trajectory of $j(z)$ deviates from $\Lambda$CDM in early phase where as it approaches to $j=1$ as $z\rightarrow -1$. Thus, jerk parameter points us the effects of interacting viscous $\Lambda(t)$ model over the $\Lambda$CDM model.
\begin{figure}[t]
\centering
\includegraphics[width=7.5cm, height=6cm]{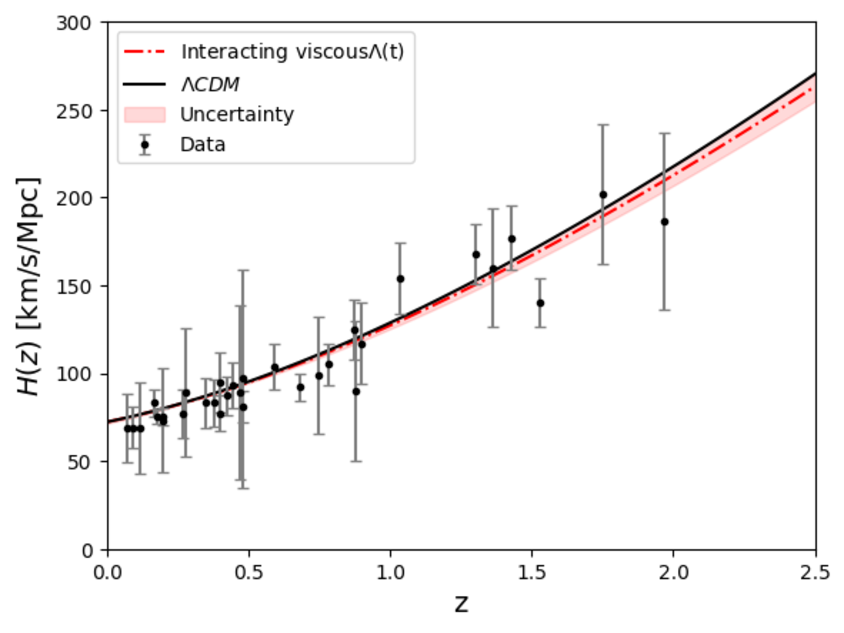}
\caption{The evolution of hubble function $H(z)$ with redshift $z$. The solid black line corresponds to the $\Lambda$CDM model and the dashed red line corresponds to the interacting viscous$\Lambda(t)$ model($\zeta= \zeta_{1} H$). The $H_{obs}(z)$ data are also plotted  with their error bars.}\label{fig:3}
\label{3}
\end{figure}

\begin{figure}[t]
\centering
\includegraphics[width=7.5cm, height=6cm]{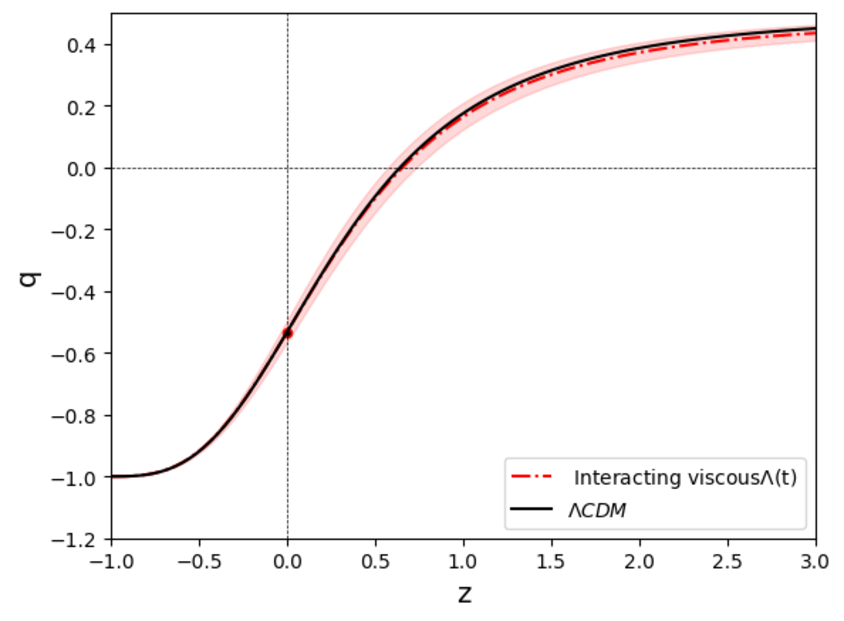}
\caption{The evolution of $q(z)$ with redshift $z$ for interacting viscous$\Lambda(t)$ model and $\Lambda$CDM model using the best fit values. Present value $q_0$ is represented by dot.}\label{fig:4}
\label{4}
\end{figure}

\begin{figure}[t]
\centering
\includegraphics[width=7.5cm, height=6cm]{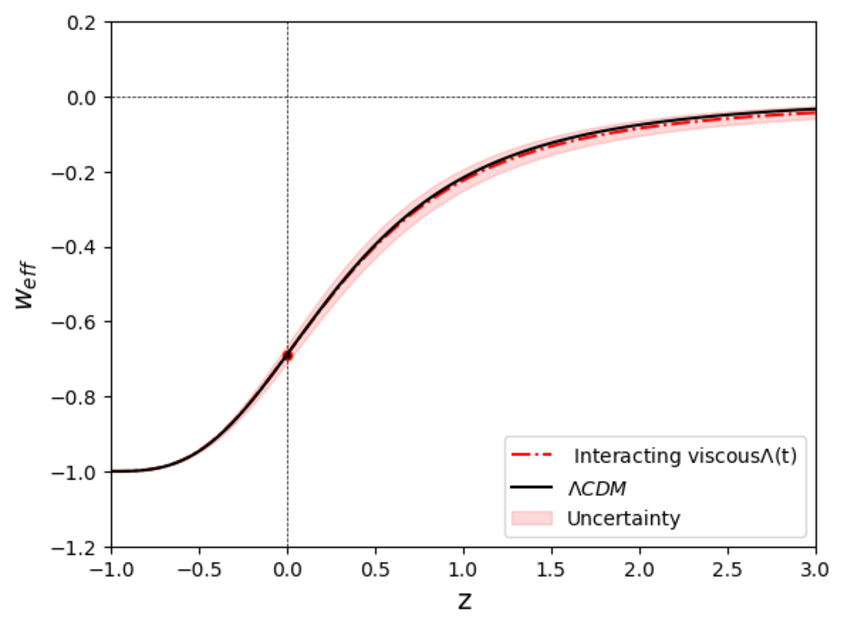}
\caption{The evolution of $w_{eff}$ with redshift $z$ for interacting viscous$\Lambda(t)$ model and $\Lambda$CDM model using the best fit values. Present value of $w_{eff}(z=0)$ is represented by dot.}\label{fig:5}
\label{5}
\end{figure}

\begin{figure}[t]
\centering
\includegraphics[width=7.5cm, height=6cm]{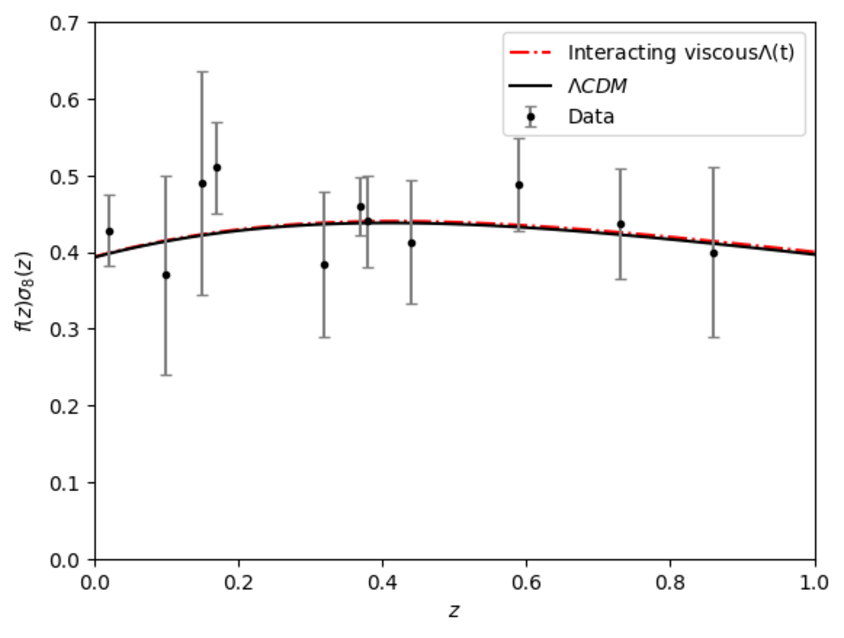}
\caption{The evolution of $f(z)\sigma_{8}(z)$ with redshift $z$ for interacting viscous$\Lambda(t)$ model and $\Lambda$CDM model using the best fit values.}\label{fig:6}
\label{6}
\end{figure}

\begin{figure}[t]
\centering
\includegraphics[width=7.5cm, height=6cm]{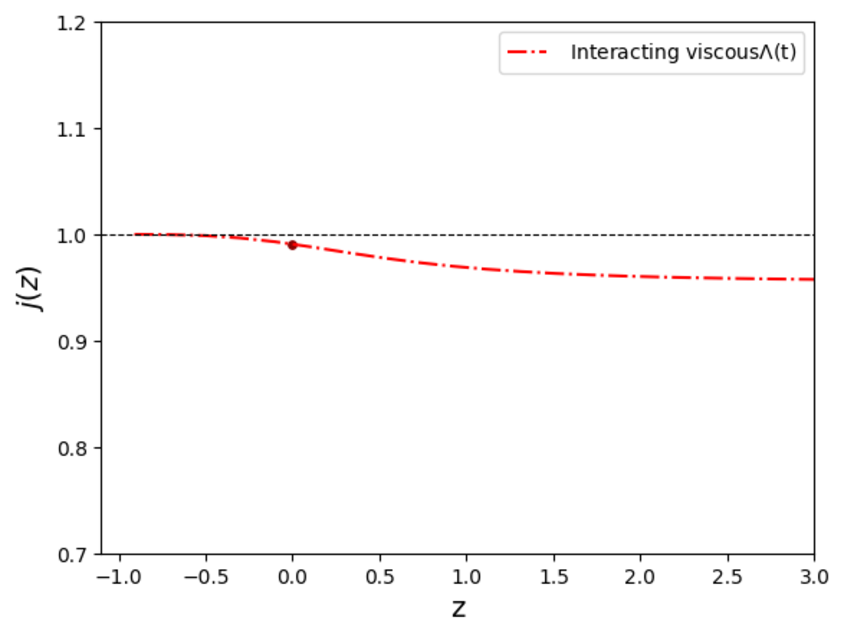}
\caption{The evolution of $j(z)$ with redshift $z$ for interacting viscous$\Lambda(t)$ model the best fit values. Present value of $j_0(z=0)$ is represented by dot and the horizontal line $j=1$ represents $\Lambda$CDM model. }\label{fig:6}
\label{7}
\end{figure}
\subsection{Model selection}\label{aicbic}
\indent In this section, we will discuss mainly reduced Chi-square and model selection criterion to observe the compatibility of the proposed model.\\
\indent Let us first calculate the reduced Chi-square, $\chi^2_{red}$, which is defined as $\chi^2_{red}=\chi^2_{min}/dof$. Here, $dof$ denotes the degree of freedom which is equal to the number of observational data points ($N$) minus the number of parameters ($K$). It is to be noted that we have used $N=1751$ and $K=3$ in $\Lambda$CDM and $K=6$ in  interacting viscous $\Lambda(t)$ model with same number of data points. Table II presents the $\chi^2$ and $\chi^2_{red}$ of $\Lambda$CDM and interacting viscous $\Lambda(t)$ models, respectively. It is noted that a value $\chi^2_{red}<1$ gives the best-fit with the data. We observed that both the models have the reduced $\chi^2$ less than unity (for $\Lambda$CDM, $\chi^2_{red}=0.447$ and for interacting viscous $\Lambda(t)$ model, it is  $\chi^2_{red}=0.448$). This shows that both the models are in very good fit with the observational data. \\
\indent We discuss the another method which provides the statistical comparison of the proposed model with $\Lambda$CDM model. In this regard, there are two model selection criterion, namely, Akaike information criteria (AIC) and Bayesian information criteria (BIC). These selection criterion allow to compare models with different degree of freedom. They are defined as \cite{akaike1974,schwarz1978}
\begin{equation}\label{aic}
AIC = \chi^{2}_{min} + \frac{2KN}{N-K-1},\;\;\;\;\;BIC = \chi^{2}_{min} + K\ln N,
\end{equation}
where $N$ is the size of the data sample and $K$ is the number of free parameters. In these approaches, the model with low values of AIC(BIC) is preferred by data. Considering the AIC (BIC) of $\Lambda$CDM model as reference model, denoted as $AIC_{\Lambda} (BIC_{\Lambda})$, we compute $\Delta AIC = AIC_{model}-AIC_{\Lambda CDM}$($\Delta BIC = BIC_{model}-BIC_{\Lambda CDM}$).\\
\indent Following the rules described in Refs.\cite{singh2024,hernandez2020}, let us discuss our results. Table IV displays the values of AIC (BIC) and corresponding $\Delta$AIC ($\Delta$BIC) for interacting viscous $\Lambda(t)$ model. According to our results, we have $\Delta AIC = 6.670$ and $\Delta BIC = 23.039$. This shows that there is ``less support in favor" of the interacting viscous $\Lambda(t)$ model as far as AIC is concerned where as BIC gives ``strong evidence against" the interacting viscous $\Lambda(t)$ model.
\subsection{Bayesian Inference}
The Bayesian evidence serves as the foundation for assessing a model's performance in light of the data. In Bayesian data analysis, the correlation between the data, the model or hypotheses, and the prior knowledge is characterized by the joint probability distributions. Given the observed data, the conditional probability distribution of the unknowns can be used to uniquely infer the posterior distribution using Bayes theorem. The key statistic of Bayesian  model selection is the Bayesian evidence $\mathcal{E}$ \cite{jeffreys1961, MacKay2003}, which is employed in a model comparison problem by integrating the product of likelihood and posterior over the entire parametric space of the model. It is defined as \cite{Liddle2006,Pia2006,Liddle2007,Trotta2007}
\begin{equation}\label{evi}
\mathcal{E}(D|M)=\int_{M} \mathcal{L}(D|\Theta, M)\mathcal{P}(D|\Theta, M)d\Theta.
\end{equation}
On the right-hand side, $\Theta$ is a set of free parameters for the given data $D$, $M$ stands for the model, the vertical bar reads as `given', and $\mathcal{ L}$ and $\mathcal{P}$ stand for likelihood and prior probability distribution function of those parameters before the data, respectively.\\
\indent In cosmology, this concept has been utilized extensively \cite{santos2017,silva2021}. When comparing two models, $M_{i}$ versus $M_{j}$, one is interested
in the ratio of the models' evidences known as $Bayes$ $factor$, which is given by:
\begin{equation}\label{ev}
\mathcal{B}_{ij} \equiv \frac{\mathcal{E}_{i}}{\mathcal{E}_{j}}
\end{equation}
Here, $B_{ij}$ indicates the support for model $i$ over model $j$.\\
\indent Bayes factors are typically interpreted using the Jeffreys' scale \cite{jeffreys1961} which measures the strength of the evidence, given in
Table III. We accomplish this by estimating the values of the logarithm of Bayes factor $(\ln\mathcal{B})$ and the Bayesian evidence $(\ln\mathcal{E})$. Using the dataset and the priors mentioned in Table I, the values of $\mathcal{E}$ and $\mathcal{B}$ are determined. We assume the $\Lambda$CDM as the reference model.\\
\indent The results for the  Bayesian evidence $(\ln\mathcal{E})$ and  Bayes factor $(\ln\mathcal{B})$ for both $\Lambda$CDM and interacting viscous $\Lambda$(t) models examined in this work are summarized in Table IV. We find that the $\ln\mathcal{B}$ for interacting viscous $\Lambda$(t) model with respect to the $\Lambda$CDM model is obtained as $2.577$ and thus the Bayesian evidence analysis shows that our model is ``moderately supported" by the considered priors and dataset on Jeffreys'scale.\\

\begin{table}
\caption{``Jeffreys' scale"  for evaluating the strength of evidence between two comparing  models, $M_{i}$ versus $M_{j}$. The right column provides the convention for denoting the different levels of evidence above these thresholds.\\}

\begin{tabular}{l l}
\hline
$\ln\mathcal{B}_{ij}$  & \;\;\;\;\;\;\;\;\;\;Strength of Evidence \\
\hline
$<1.0$ & \;\;\;\;\;\;\;\;\;\;Inconclusive\\

$1.0$ & \;\;\;\;\;\;\;\;\;\;Weak evidence\\

$2.5$ & \;\;\;\;\;\;\;\;\;\;Moderate evidence\\

$5.0$ & \;\;\;\;\;\;\;\;\;\;Strong evidence\\

\hline
\end{tabular}
\label{table:jeffrey}
\end{table}
\begin{table}
\caption{Summary of $\ln\mathcal{E}$, $\ln\mathcal{B}$, $\chi^2$, $\chi^2_{red}$, $AIC$  and $BIC$ for the $\Lambda$CDM and interacting viscous$\Lambda(t)$ models for the following:}

\begin{tabular}{ |l||c|c|}

%\multicolumn{4}{|c|}{Viscous $\Lambda$(t)} \\
\hline
Values & $\Lambda$CDM & Interacting viscous$\Lambda(t)$  \\
\hline
$\ln\mathcal{E}$ & $-793.120 \pm 0.473$ & $-795.697 \pm 0.532 $\\

$\ln\mathcal{B}$ & $-$ & $2.577$ \\

$\chi^2$  & $781.708$ & $782.344$   \\

$K$ & $3$ & $6$  \\

$N$ & $1751$ & $1751$  \\

$\chi^2_{red}$ & $0.447$ & $0.448$   \\

$AIC$ & $787.722$ & $794.392$ \\	

$\Delta AIC$  & $-$ & $6.670$  \\

$BIC$ & $804.112$ & $827.152$    \\

$\Delta BIC$  & $-$ & $23.039$ \\

\hline
\end{tabular}
\label{table:values1}
\end{table}
\section{Conclusion}\label{conclusion}
Inspired by dissipative phenomena and decaying vacuum energy, we have discussed some cosmological consequences of an alternative mechanism of accelerating Universe based on a class of interacting viscous model with decaying vacuum energy, referred as interacting viscous $\Lambda(t)$ model. The coupling between viscous fluid and vacuum energy density has been made through a coupling parameter, $Q$. Within the framework of Eckart thermodynamic theory the non-equilibrium pressure, $\Pi$ is proportional to the Hubble parameter $H$ with proportionality constant $\zeta$, i.e., $\Pi=-3\zeta (t) H$.  Thus, the effective pressure is assumed as sum of barotropic pressure, bulk viscous pressure and pressure due to vacuum energy, i.e., $\bar{p}=p_m+\Pi+p_{\Lambda}$. In the first part of work, we have obtained some main results for the scale factor, the Hubble parameter, the deceleration parameter, jerk parameter and EoS parameter by assuming the interaction term $Q=3\alpha \rho_m H$. Although the nature of bulk viscosity and time-varying vacuum energy density are unknown, we have assumed $\zeta=\zeta_1H$ for bulk viscous coefficient and $\rho_{\Lambda}=c_0+3\nu H^2$ for vacuum energy density. One can expect that the viscosity is affected by the expansion rate of the Universe and varying VED from the general covariance of the effective action in QFT. We have investigated the growth perturbation of interacting viscous $\Lambda(t)$ model. In the second part of this work, we have performed the Bayesian analysis using the latest background probes such as SNe Pantheon+, cosmic chronometer and $f(z)\sigma_8(z)$. We have compared the interacting viscous $\Lambda(t)$ model with $\Lambda$CDM using the Bayesian inference and model selection criterion such as AIC and BIC. In what follows, we summarize the main points of our analysis.\\
\indent The constraints on model free parameters have been reported in Table II. The best-fit value of $H_0$ according to a combination of Pantheon+, CC and $f(z)\sigma_8(z)$ is $H_0 = 72.202^{+0.796}_{-0.937}$ $km s^{-1} Mpc^{-1}$, which is tension with both Planck and R21 results at $4.80\sigma$ and $0.86\sigma$ ,respectively. In other words, the tension in $H_0$ measurement is almost resolved in interacting viscous $\Lambda(t)$ model with respect to R21. The best fit values of present $q$ and $w_{eff}$ are $q_0=-0.531^{+0.024}_{-0.024}$ and $w_\text{eff}(z=0)=-0.687\pm 0.016$, respectively. From Fig.5, it has been observed that the interacting viscous $\Lambda(t)$ model exhibits transition from an early decelerated phase to late-time accelerated phase and the transition takes place at $z_{tr} =0.647^{+0.038}_{-0.044}$ which is close proximity $z_{tr}= 0.643^{+0.041}_{-0.041}$ to that of $\Lambda$CDM model. It has been found that the value of $\chi^2_{red}<1$ which shows that the interacting viscous $\Lambda(t)$ model is in very good fit with the used data points. The jerk parameter remains positive and less than one in past, and tends to unity in late-time.\\
\indent We have explored the $\sigma_8$ and $S_8$ parameters for $\Lambda$CDM and interacting viscous $\Lambda(t)$ models using the combined data set of Pantheon+, CC and $f(z)\sigma_8(z)$. The constraints on $\sigma_8$ and $S_8$ in interacting viscous $\Lambda(t)$ model are $\sigma_8 = 0.739^{+0.034}_{-0.030}$ and $S_8 = 0.762^{+0.030}_{-0.030} $ which are very close to $\Lambda$CDM model as reported in Table II. The tensions of our fitting results in $\sigma_8$ and $S_8$ with respect to Plank results \cite{plank2020} are $2.21\sigma$ and $2.11\sigma$, respectively. The evolution of $f(z)\sigma_8(z)$ has been plotted in Fig. 7 which shows that it is consistent with the observational data points. \\
\indent It is noted that cosmological models are testable from the abundance of observational data as discussed above. However, an important distinction must be made between parameter fitting and model selection. As we know that the parameter fitting simply tells us how well a model fit the data. Model selection such as Bayesian inference and  AIC and BIC are necessary to discriminate the proposed model with the existing model. The Bayesian inference analysis demonstrated the interacting viscous $\Lambda$(t) model is moderately supported by the considered dataset and priors. Further, there has been increasing interest in applying information criterion such as AIC and BIC for model selection. We have examined the models using AIC and BIC for a fairer comparison. We have observed that the interacting viscous $\Lambda(t)$ model has ``less support"  according to the selection criteria $\Delta$AIC. On contrary, with respect to $\Delta$BIC, interacting viscous $\Lambda(t)$ model has ``strong evidence against" the model with the considered datasets. \\
\indent As a concluding remark we must point out that despite its intrinsic nature of bulk viscosity and decaying vacuum energy and its interaction are not well understood yet, the work presented in this paper suggests a possible description for resolving the $H_0$ and $\sigma_8$ tensions in cosmology. In principle, to give a robust approach for investigating the dark energy model beyond $\Lambda$CDM, background dynamics should be considered. Taking into account such interaction between the dark components may provide an opportunity to explain the present accelerating Universe. Indeed considering the interaction between the viscous fluid and decaying vacuum energy potentially enables us to resolve tensions in cosmological parameters.

\begin{acknowledgements}
One of the author, VK would like to express gratitude to Delhi Technological University, India for providing Research Fellowship to carry out this work.
\end{acknowledgements}

\section{Appendix}
The main aim in this appendix is to present some more analytical solutions of the cosmological parameters of the interacting viscous model with decaying vacuum energy density based on the different choices of $\zeta$.
\appendix
\section{Solution with $\zeta=\zeta_0$ }
It is the most simple parametric form of the bulk viscous coefficient. Many authors \cite{brevik2005,hu2006,avelino2009,singh2018,singh2019,singh2020,nour2011,ajay2019,chitre1987,montiel2011} have studied the viscous cosmological models with constant bulk viscous coefficient. Using $\zeta=\zeta_0$ in Eq.\eqref{eqh}, the Hubble evolution equation deduces the form
\begin{equation}\label{eq13}
\dot{H}+\frac{3}{2}(1-\alpha)H^2-\frac{3}{2} \frac{\zeta_0 H}{(1-\nu)} H = \frac{1}{2}\left(\frac{1-\alpha}{1- \nu}\right)c_0.
\end{equation}
\indent Solving \eqref{eq13} with the condition $H(t_0)=H_0$, we get
\begin{equation}\label{eq15}
H=\frac{\zeta_0}{2(1-\nu)(1-\alpha)}+ k \coth\left(\frac{3}{2}(1-\alpha)k t\right).
\end{equation}
\indent where $k = \frac{{\zeta_0}^2 + 4(\Omega_{\Lambda} - \nu)(1-\nu){(1-\alpha)}^2{H_0}^2}{{4(1 - \nu)}^2{(1-\alpha)}^2}$.\\
Integrating again with the condition $a(t_0)=1$, we obtain the scale factor
\begin{equation}\label{eq16}
a(t)=e^{\frac{\zeta_0}{2(1-\nu)(1-\alpha)}t}\left[\sinh \left(\frac{3}{2}(1-\alpha)k t\right)\right]^{\frac{2}{3(1-\alpha)}},
\end{equation}
Using \eqref{eq15}, the deceleration parameter $q$ and Effective EoS parameter $w_{eff}$ are respectively given by
\begin{equation}\label{eq18}
q=-1+\frac{3}{2}\frac{(1-\alpha)k^2 \csc^2h (\frac{3}{2}(1-\alpha)k t)}{\left(\frac{\zeta_0}{2(1-\nu)(1-\alpha)}+ k\coth(\frac{3}{2}(1-\alpha)k t)\right)^2}.
\end{equation}
\begin{equation}\label{eq20}
w_{eff}=-1+\frac{(1-\alpha)k^2 \csc^2h (\frac{3}{2}(1-\alpha)k t)}{\left(\frac{\zeta_0}{2(1-\nu)(1-\alpha)}+k\coth(\frac{3}{2}(1-\alpha)k t)\right)^2}.
\end{equation}
\section{Solution with $\zeta=\zeta_0+\zeta_1 H$}
This form of bulk viscous coefficient has been discussed by several authors and the references therein \cite{meng2007,meng2009,avelino2010}. For $\zeta=\zeta_0+\zeta_1 H$, where $\zeta_0$ and $\zeta_1$ are constants, the Eq. \eqref{eqh} simplifies into
\begin{equation}\label{eqc1}
\dot{H}+\frac{3}{2}(1-\alpha)H^2-\frac{3}{2} \frac{\zeta_1 H^2}{(1-\nu)} -\frac{3}{2} \frac{\zeta_0 H}{(1-\nu)} = \frac{1}{2}\left(\frac{1-\alpha}{1- \nu}\right)c_0,
\end{equation}
which can be integrated to calculate the Hubble function and the scale factor

\small
\begin{equation}\label{eqc3}
H=\frac{\zeta_0}{2((1-\alpha)(1-\nu)-\zeta_1)}+k_1\; \coth\left(\frac{3}{2}\frac{((1-\alpha)(1-\nu)-\zeta_1)}{(1-\nu)}k_1 t\right),
\end{equation}

\begin{eqnarray}\label{eqc4}
a&=&e^{\frac{\zeta_0\;t}{2((1-\alpha)(1-\nu)-\zeta_1)}}\nonumber\\
& & \times \left[\sinh\left(\frac{3}{2}\frac{((1-\alpha)(1-\nu)-\zeta_1)}{1-\nu}k_1t\right)\right]^{\frac{2(1-\nu)}{3((1-\alpha)(1-\nu)-\zeta_1)}},
\end{eqnarray}
%\end{widetext}
where $k_1 = \frac{{\zeta_0}^2 + 4((1-\alpha)(1-\nu)-\zeta_1)((\Omega_{\Lambda} - \nu)(1-\alpha)){H_0}^2}{4{((1 - \nu)(1-\alpha)-\zeta_1)}^2}$. \\

\indent With this, we can calculate the deceleration parameter $q$ and effective equation of state parameter $w_{eff}$, which are respectively given by
\begin{equation}\label{eqc5}
q=-1
 +\frac{3(1-\zeta_1-\nu)\sigma^2_1 \csc^2h (\frac{3}{2}(1-\zeta_1-\nu)\sigma_1 t)}{2\left(\frac{\zeta_0}{2(1-\zeta_1-\nu)}+\sigma_1 \coth(\frac{3}{2}(1-\zeta_1-\nu)\sigma_1 t)\right)^2}
\end{equation}
and\\
\begin{equation}\label{eqc6}
w_{eff}=-1+\frac{(1-\zeta_1-\nu)\sigma^2_1 \csc^2h (\frac{3}{2}(1-\zeta_1-\nu)\sigma_1 t)}{\left(\frac{\zeta_0}{2(1-\zeta_1-\nu)}+\sigma_1 \coth(\frac{3}{2}(1-\zeta_1-\nu)\sigma_1 t)\right)^2}
\end{equation}

\end{document}